\documentstyle[prb,multicol,aps,epsfig]{revtex}

\begin{document}

\title{Combining high conductivity with complete 
optical transparency: \\ A band-structure approach}

\author{J.~E. Medvedeva and A.~J. Freeman}

\address{
Department of Physics and Astronomy, Northwestern,
Evanston, Illinois 60208-3112
}

\maketitle

\begin{abstract}
A comparison of the structural, optical and electronic properties
of the recently discovered transparent conducting oxide (TCO), 
nanoporous Ca$_{12}$Al$_{14}$O$_{33}$, with those of the conventional
TCO's (such as Sc-doped CdO) indicates that this material belongs 
conceptually to a new class of transparent conductors.
For this class of materials, we formulate criteria for the
successful combination of high electrical conductivity with 
{\it complete} transparency in the visible range.
Our analysis suggests that this set of requirements can be met
for a group of novel materials called electrides.
\end{abstract}

{\bf PACS number(s)}:
71.20.-b, 
72.20.-i, 
78.20.Bh 

\begin{multicols}{2}

Transparent conducting oxides (TCO) have been known for almost
a century \cite{Badeker} and employed technologically for decades. 
Today, the area of practical applications of this special class 
of materials which can simultaneously act as a window layer and 
as an electrically 
conducting contact layer, is very large \cite{Thomas,MRS,Stoute}: 
it includes optoelectronics (invisible circuits), flat-panel displays, 
energy supply (solar cells) and energy conservation 
(``smart'' windows) devices.
The commercial demand for less expensive, more flexible, 
environmentally friendly materials that exhibit both high optical 
transmission and electrical 
conductivity continues to stimulate further research.

All well-known and widely used TCO's (such as In, Sn, Zn, 
Cd, Ga and Cu oxides and their blends) share similar
chemical, structural and electronic properties as well as 
carrier generation mechanisms. 
These oxides of post-transition (or transition) metals have 
relatively close-packed structure with four or six-fold coordinated 
metal ions. Upon introduction of native or substitutional dopants,
they show high transparency in the visible range ($\sim$80-90 \%) and 
high electrical conductivity (up to $\sim$10$^4$ S/cm).
Common to all known TCO's, a highly dispersed band 
at the bottom of the conduction bands is the most important feature 
of the host electronic band structure. It provides both (i) the high 
mobility of the extra carriers (electrons) due to their small effective 
masses and (ii) low optical absorption due to a pronounced Burstein-Moss 
shift which helps to keep intense interband transitions out of 
the visible range \cite{Mryasov}.

\begin{figure}
\includegraphics[width=4.1cm]{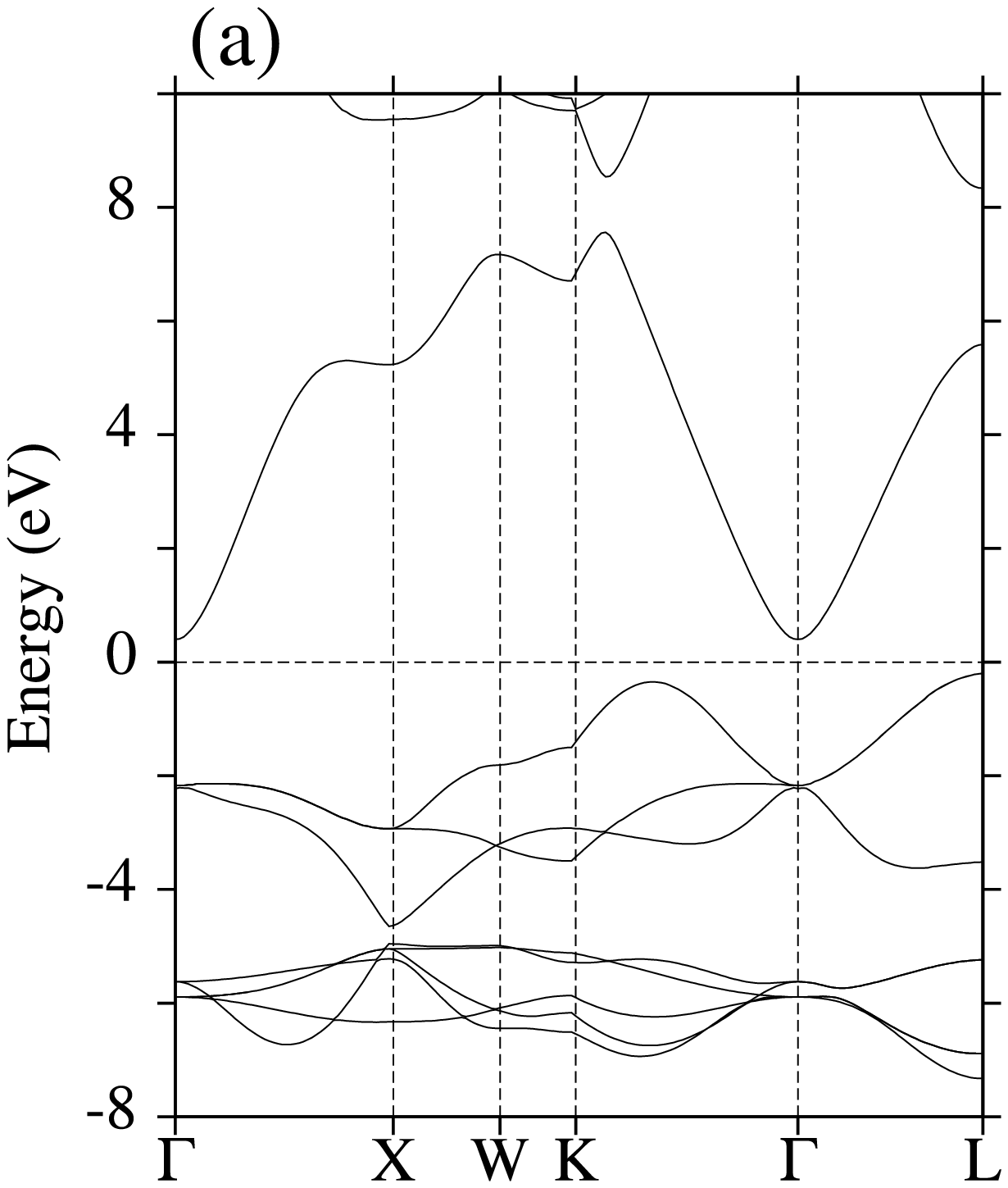}
\includegraphics[width=4.1cm]{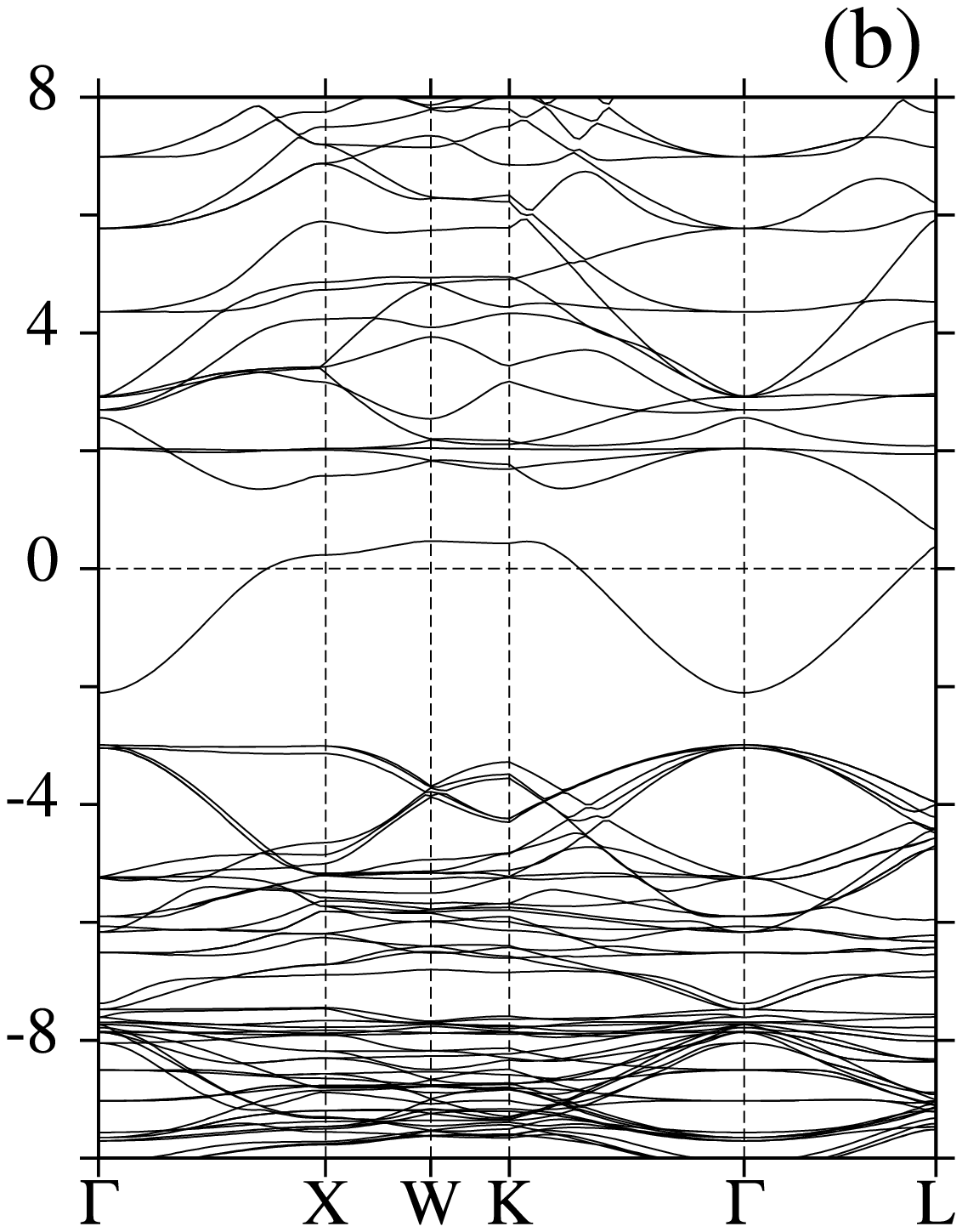}
\caption{The electronic band structure of (a) undoped 
and (b) Sc-doped CdO.}
\label{cdo}
\end{figure}

To illustrate how doping alters the electronic band structure of host 
transparent conductors, we present the calculated band structure 
of undoped and Sc-doped CdO, cf., Fig. \ref{cdo}, determined with
the full-potential linearized augmented plane wave method \cite{FLAPW} 
(FLAPW) within the screened-exchange LDA approach \cite{sxLDA}. 
Despite a rather small (indirect) bandgap of $\sim$1 eV in pure CdO 
(as compared to bandgaps of $\sim$3.0 eV in undoped In$_2$O$_3$, 
ZnO and SnO$_2$), the optical window significantly broadens upon
doping (Burstein-Moss shift) so that the intense interband transitions
from the valence band are $>$3.0 eV (Fig. \ref{cdo}(b)). 
The high dispersion of the band at the bottom of the conduction band 
also provides a relatively low intensity of interband transitions 
from the partially occupied band at the top of the conduction band.
Finally, the small effective masses obtained in pure CdO, 0.25 m$_e$, 
contribute to the observed high carrier mobility in doped CdO 
(up to $\sim$600 cm$^2$/V~s, Ref. \onlinecite{Chang}).

Recently, an insulator-conductor conversion was discovered
in an oxide that differs essentially from the known TCO by its chemical 
and structural properties and by the origin of the induced conductivity: 
the cage structured 
insulating calcium-aluminum oxide, 12CaO$\cdot$7Al$_2$O$_3$, or mayenite,
showed a persistent conductivity upon doping with hydrogen 
followed by ultraviolet irradiation \cite{Hayashi2002}.
First-principles electronic band structure calculations \cite{origin}
have already revealed that the charge transport associated with 
the electrons excited off the hydrogen ions (H$^-$$\rightarrow$H$^0$+e$^-$) 
occurs by electron hopping 
through the encaged ``defects'' -- the H$^0$ and OH$^-$ located 
inside the large (more than 5.6 \AA \, in diameter) structural cavities.
The low conductivity of the material ($\sim$1 S/cm) was attributed 
to the strong interactions between the UV released electrons which 
migrate along a narrow conducting channel -- the hopping path.
Indeed, the alleviation of their electronic repulsion \cite{electride} resulted in 
the observed \cite{Matsuishi2003} 100-fold enhancement of the conductivity 
in the mayenite-based oxide, [Ca$_{12}$Al$_{14}$O$_{32}$]$^{2+}$(2e$^-$),
although the carrier concentration is only two times larger than that
in the H-doped UV-irradiated Ca$_{12}$Al$_{14}$O$_{33}$.
The improved conductivity, however, came at the cost of greatly increased
absorption \cite{electride,Matsuishi2003}, making this oxide unsuitable 
for practical use as a transparent conducting material. 

\begin{figure}
\includegraphics[width=4.2cm]{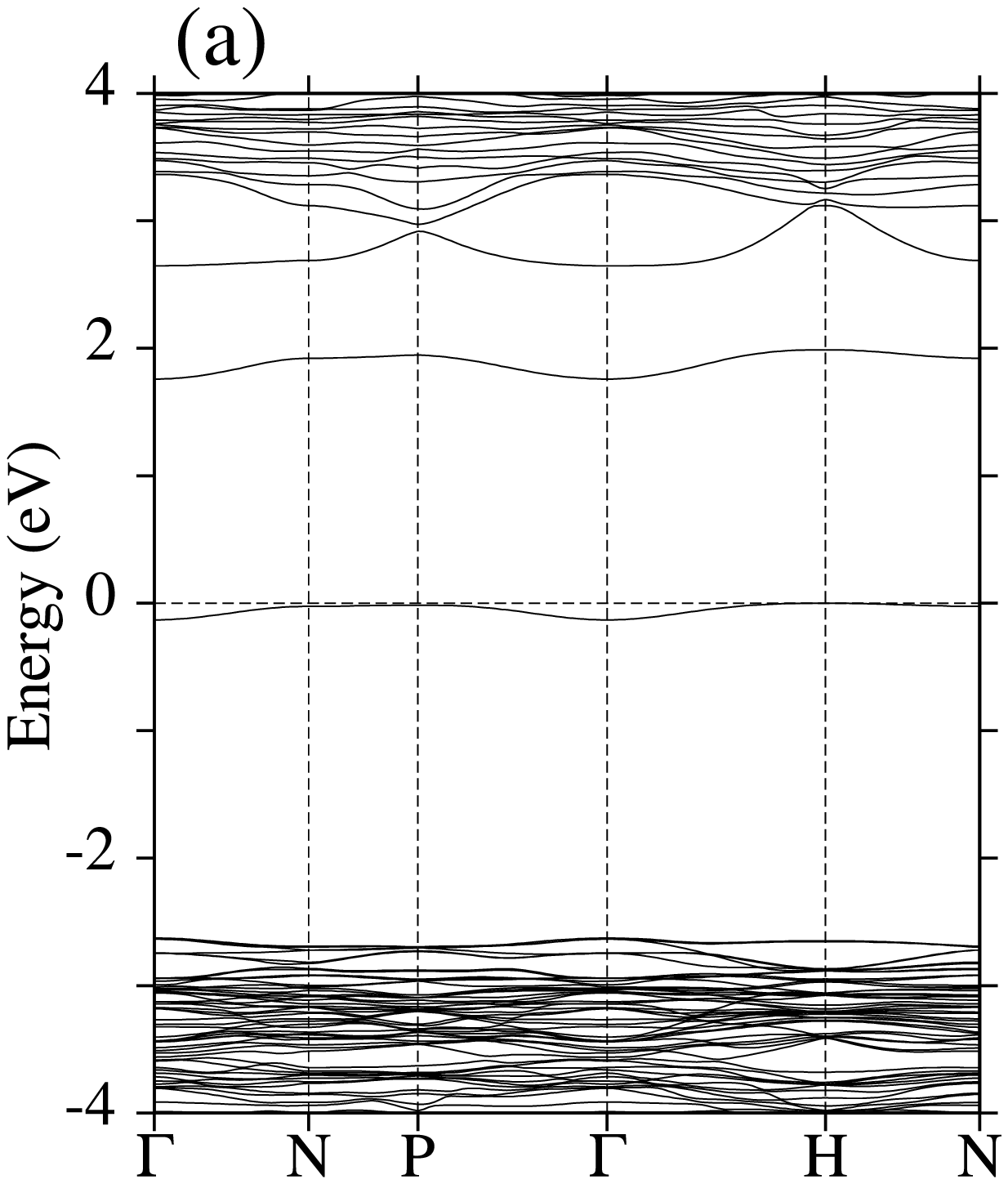}
\includegraphics[width=4.2cm]{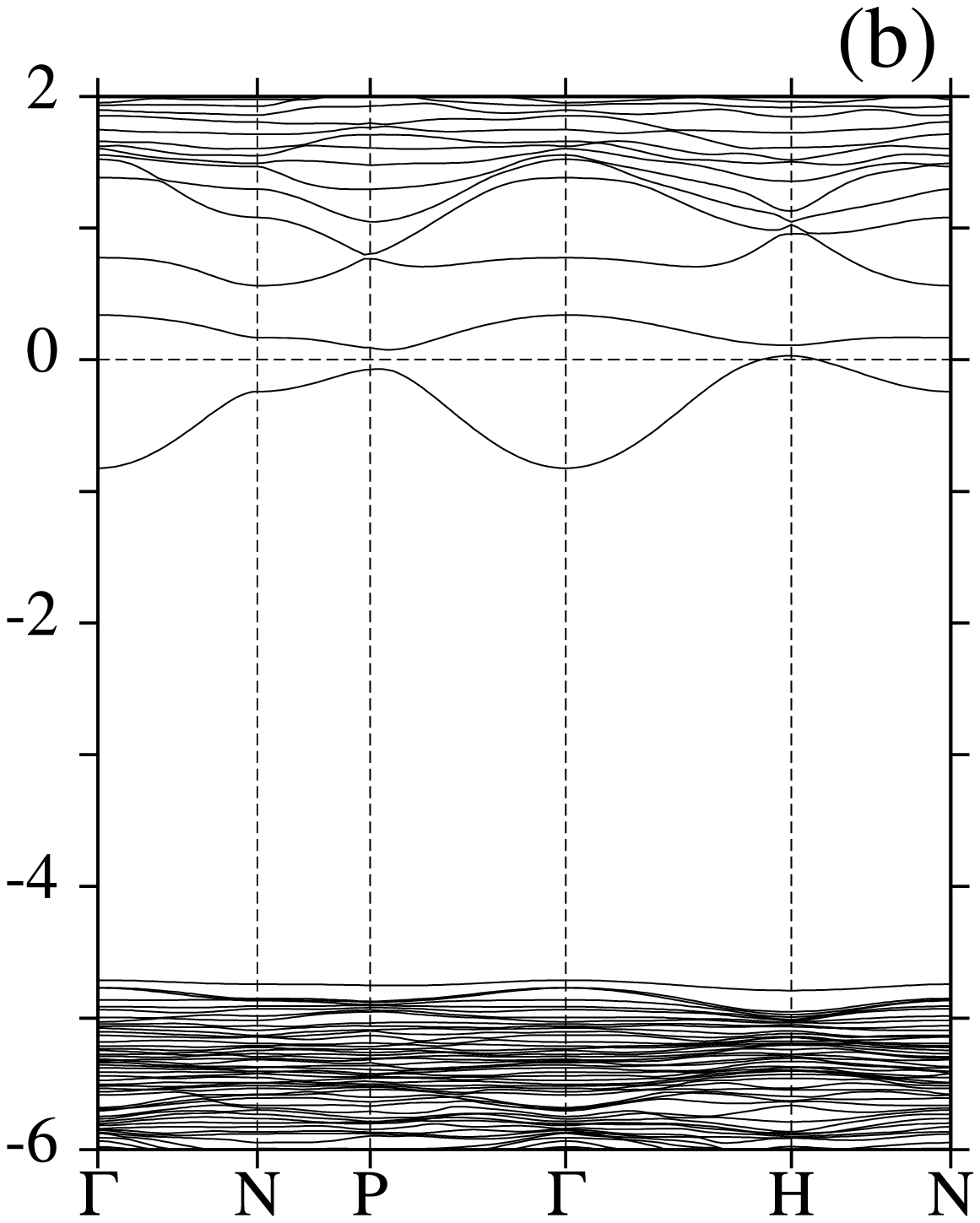}
\caption{The electronic band structure of H-doped mayenite
(a) before and (b) after UV-irradiation.
}
\label{may}
\end{figure}

Despite the failure to combine effectively the low optical transparency 
and useful electrical conductivity in the mayenite-based oxides, 
the nature of their electronic band structure \cite{origin,electride} 
suggests that these materials belong to a conceptually new class of TCO's.
As shown in Fig. \ref{may}, hydrogen annealing and the subsequent 
UV-irradiation of Ca$_{12}$Al$_{14}$O$_{33}$ results in the formation
of a {\it new} hybrid ``defect'' band in the band gap of insulating mayenite. 
This band crosses the Fermi level making the system conducting.
Further, the transitions from the occupied part of the band to 
the unoccupied one are below the visible range due to the
narrowness of the hybrid band, while the interband transitions 
to the bottom of the conduction band are rather weak since they are provided by 
the low density of states (DOS) in the hybrid band near E$_F$.
Consequently, any increase of the DOS at E$_F$ that favors a higher
conductivity, would result in an increase of light absorption, 
reducing the optical transmission. 
Indeed, this was observed \cite{Matsuishi2003} 
for [Ca$_{12}$Al$_{14}$O$_{32}$]$^{2+}$(2e$^-$), where the DOS
at E$_F$ is found to be 24 times larger than in
the H-doped UV-irradiated mayenite \cite{electride}.

\begin{figure}
\centerline{
\includegraphics[width=5.8cm]{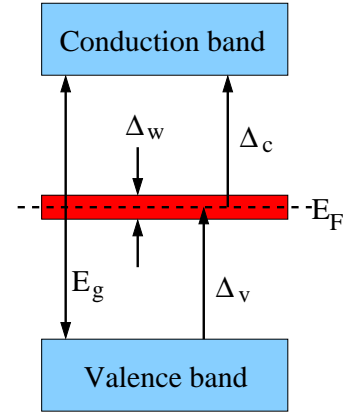}
}
\caption{The schematic band structure of ``ideal'' TCO: 
$\Delta _w<$ 1.8 eV, $\Delta _v>$ 3.1 eV and $\Delta _c>$ 3.1 eV
provide 100\% transparency in the visible range.}
\label{scheme}
\end{figure}

Thus, in striking contrast to the conventional TCO's, where the optical 
absorption cannot be eliminated, the band structure analysis of 
mayenite-based oxides suggests an approach to combine 100\% optical 
transparency and high electrical conductivity.
The schematic band structure of such an ``ideal'' TCO is shown
in Fig. \ref{scheme}. 
The introduction of a {\it deep} impurity band in the bandgap of an insulating 
material would help to keep intense interband transitions (from the valence
band to the impurity band and from the impurity band to the conduction band)
above the visible range. 
This requires the band gap of a host material to be more than 6.2 eV. 
Furthermore, the impurity band should be narrow enough (less than 1.8 eV) 
to keep intraband transitions (as well as the plasma frequency) 
below the visible range.

To achieve high conductivity, the concentration of impurities should be 
large enough so that their electronic wavefunctions overlap and form
an impurity {\it band}. The formation of the band would lead to a high 
carrier mobility due to the extended nature of these states 
resulting in a relatively low scattering.
For this, a material with a close-packed structure should not be used, 
because large concentration of impurities would result in
(i) an increase of ionized impurity scattering which limits 
electron transport \cite{BrooksDingle}; and (ii) large 
relaxation of the host material, affecting its electronic structure and, 
most likely, decreasing the desired optical transparency.
Alternatively, nanoporous structure materials offer a way to incorporate a
large concenration of impurities without any significant changes in
the band structure of the host material, e.g., H$^-$ and OH$^-$ in the 
spacious cages of mayenite. 
However, such encaged impurities would be well-separated from each other
and, therefore, would not form by themselves an impurity band that is necessary 
for creating extended well-conducting electronic states \cite{cagewall}.
The coupling between impurities can be achieved by choosing a material 
where an array of connected structural cavities (channels) will allow 
the interaction of the nearby impurities -- unlike the mayenite-type 
materials where the encaged impurities are isolated. 
For this, novel materials called electrides \cite{Dye90-93}
seem to be very promising candidates 
for ``ideal'' TCO's due to their unique structural, optical and
electronic properties -- namely, intercavity channels \cite{note}, 
large bandgaps, weak binding of the ``excess'' electrons and 
near-metal electronic conductivity \cite{Dye,Singh,Petkov,Li}.

Finally, while the conductivity in the proposed new type of TCO materials 
may not exceed the maximum value of the conventional TCO's, 
i.e., $\sim$10$^4$ S/cm (due to similar intrinsic limits \cite{limits}),
their optical transparency can be expected to be as high as 100\% 
in the visible range of wavelength.

Work supported by the DOE (grant N DE-FG02-88ER45372)
and computational resources provided by NERSC.


\end{multicols}

\begin{thebibliography}{99}
\bibitem{cagewall}
In mayenite, this was successfully overcome by the UV-irradiation which
resulted in the formation of the hybrid band. However, the band was formed 
not only from the states of the encaged impurities but also from the states 
of atoms belonging to the cage wall (namely, 3d states of Ca atoms
which form the bottom of the conduction band).
This led to degradation of the optical properties since the hybrid band was 
effectively shifted towards the conduction band minimum.
\end{thebibliography}
\end{document}